\newif\ifprl

\prltrue

\ifprl
\documentclass[aps,prl,superscriptaddress,twocolumn,showpics]{revtex4}

\else
\input amssym.def
\input amssym.tex
\fi

\newcommand{\beq}{\begin{equation}}
\newcommand{\eeq}{\end{equation}}
\newcommand{\bbf}{\begin{bfseries}}
\newcommand{\ebf}{\end{bfseries}}
\newcommand{\bbm}{\begin{boldmath}}
\newcommand{\ebm}{\end{boldmath}}
\newcommand{\beqn}{\begin{eqnarray}}
\newcommand{\eeqn}{\end{eqnarray}}
\newcommand{\bi}{\begin{itemize}}
\newcommand{\ei}{\end{itemize}}
\newcommand{\bd}{\begin{description}}
\newcommand{\ed}{\end{description}}
\newcommand{\ben}{\begin{enumerate}}
\newcommand{\een}{\end{enumerate}}
\newcommand{\bHuge}{\begin{Huge}}
\newcommand{\bhuge}{\begin{huge}}
\newcommand{\bLARGE}{\begin{LARGE}}
\newcommand{\bLarge}{\begin{Large}}
\newcommand{\blarge}{\begin{large}}
\newcommand{\eHuge}{\end{Huge}}
\newcommand{\ehuge}{\end{huge}}
\newcommand{\eLARGE}{\end{LARGE}}
\newcommand{\eLarge}{\end{Large}}
\newcommand{\elarge}{\end{large}}

\newcommand{\ra}{\rightarrow}
\newcommand{\piz}{\mbox{$\pi^0$}}

\newcommand{\met}{\mbox{${E\!\!\!\!/_T}$}}

\newcommand{ \hpm}{\mbox{$H^{\pm \pm}$}}
\newcommand{ \hpml}{\mbox{$H_L^{\pm \pm}$}}
\newcommand{ \hpmr}{\mbox{$H_R^{\pm \pm}$}}

\newcommand{\gev}  {\mbox{$\;{\rm GeV}$}}

\newcommand{\gevc} {\mbox{$\;{\rm GeV}/c$}}
\newcommand{\gevcc}{\mbox{$\;{\rm GeV}/c^2$}}

\newcommand{\shugelist}{\begin{center}\begin{Huge}\begin{itemize}}
\newcommand{\ehugelist}{\end{itemize}\end{Huge}\end{center}}
\newcommand{\bc}{\begin{center}}
\newcommand{\ec}{\end{center}}

\usepackage{multirow}
 \usepackage{graphicx,thumbpdf}
\ifprl

\else
\fi

\begin{document}

\ifprl
\bibliographystyle{apsrev}
\else
\doublespace
\bibliographystyle{prsty}
\fi

\ifprl
\preprint{}
\fi
\ifprl
\title{Search for Doubly Charged Higgs Bosons with Lepton-Flavor-Violating Decays involving Tau Leptons}
\fi

\affiliation{Institute of Physics, Academia Sinica, Taipei, Taiwan 11529, Republic of China} 
\affiliation{Argonne National Laboratory, Argonne, Illinois 60439} 
\affiliation{Institut de Fisica d'Altes Energies, Universitat Autonoma de Barcelona, E-08193, Bellaterra (Barcelona), Spain} 
\affiliation{Baylor University, Waco, Texas  76798} 
\affiliation{Istituto Nazionale di Fisica Nucleare, University of Bologna, I-40127 Bologna, Italy} 
\affiliation{Brandeis University, Waltham, Massachusetts 02254} 
\affiliation{University of California, Davis, Davis, California  95616} 
\affiliation{University of California, Los Angeles, Los Angeles, California  90024} 
\affiliation{University of California, San Diego, La Jolla, California  92093} 
\affiliation{University of California, Santa Barbara, Santa Barbara, California 93106} 
\affiliation{Instituto de Fisica de Cantabria, CSIC-University of Cantabria, 39005 Santander, Spain} 
\affiliation{Carnegie Mellon University, Pittsburgh, PA  15213} 
\affiliation{Enrico Fermi Institute, University of Chicago, Chicago, Illinois 60637} 
\affiliation{Comenius University, 842 48 Bratislava, Slovakia; Institute of Experimental Physics, 040 01 Kosice, Slovakia} 
\affiliation{Joint Institute for Nuclear Research, RU-141980 Dubna, Russia} 
\affiliation{Duke University, Durham, North Carolina  27708} 
\affiliation{Fermi National Accelerator Laboratory, Batavia, Illinois 60510} 
\affiliation{University of Florida, Gainesville, Florida  32611} 
\affiliation{Laboratori Nazionali di Frascati, Istituto Nazionale di Fisica Nucleare, I-00044 Frascati, Italy} 
\affiliation{University of Geneva, CH-1211 Geneva 4, Switzerland} 
\affiliation{Glasgow University, Glasgow G12 8QQ, United Kingdom} 
\affiliation{Harvard University, Cambridge, Massachusetts 02138} 
\affiliation{Division of High Energy Physics, Department of Physics, University of Helsinki and Helsinki Institute of Physics, FIN-00014, Helsinki, Finland} 
\affiliation{University of Illinois, Urbana, Illinois 61801} 
\affiliation{The Johns Hopkins University, Baltimore, Maryland 21218} 
\affiliation{Institut f\"{u}r Experimentelle Kernphysik, Universit\"{a}t Karlsruhe, 76128 Karlsruhe, Germany} 
\affiliation{Center for High Energy Physics: Kyungpook National University, Daegu 702-701, Korea; Seoul National University, Seoul 151-742, Korea; Sungkyunkwan University, Suwon 440-746, Korea; Korea Institute of Science and Technology Information, Daejeon, 305-806, Korea; Chonnam National University, Gwangju, 500-757, Korea} 
\affiliation{Ernest Orlando Lawrence Berkeley National Laboratory, Berkeley, California 94720} 
\affiliation{University of Liverpool, Liverpool L69 7ZE, United Kingdom} 
\affiliation{University College London, London WC1E 6BT, United Kingdom} 
\affiliation{Centro de Investigaciones Energeticas Medioambientales y Tecnologicas, E-28040 Madrid, Spain} 
\affiliation{Massachusetts Institute of Technology, Cambridge, Massachusetts  02139} 
\affiliation{Institute of Particle Physics: McGill University, Montr\'{e}al, Canada H3A~2T8; and University of Toronto, Toronto, Canada M5S~1A7} 
\affiliation{University of Michigan, Ann Arbor, Michigan 48109} 
\affiliation{Michigan State University, East Lansing, Michigan  48824} 
\affiliation{University of New Mexico, Albuquerque, New Mexico 87131} 
\affiliation{Northwestern University, Evanston, Illinois  60208} 
\affiliation{The Ohio State University, Columbus, Ohio  43210} 
\affiliation{Okayama University, Okayama 700-8530, Japan} 
\affiliation{Osaka City University, Osaka 588, Japan} 
\affiliation{University of Oxford, Oxford OX1 3RH, United Kingdom} 
\affiliation{University of Padova, Istituto Nazionale di Fisica Nucleare, Sezione di Padova-Trento, I-35131 Padova, Italy} 
\affiliation{LPNHE, Universite Pierre et Marie Curie/IN2P3-CNRS, UMR7585, Paris, F-75252 France} 
\affiliation{University of Pennsylvania, Philadelphia, Pennsylvania 19104} 
\affiliation{Istituto Nazionale di Fisica Nucleare Pisa, Universities of Pisa, Siena and Scuola Normale Superiore, I-56127 Pisa, Italy} 
\affiliation{University of Pittsburgh, Pittsburgh, Pennsylvania 15260} 
\affiliation{Purdue University, West Lafayette, Indiana 47907} 
\affiliation{University of Rochester, Rochester, New York 14627} 
\affiliation{The Rockefeller University, New York, New York 10021} 
\affiliation{Istituto Nazionale di Fisica Nucleare, Sezione di Roma 1, University of Rome ``La Sapienza," I-00185 Roma, Italy} 
\affiliation{Rutgers University, Piscataway, New Jersey 08855} 
\affiliation{Texas A\&M University, College Station, Texas 77843} 
\affiliation{Istituto Nazionale di Fisica Nucleare, University of Trieste/\ Udine, Italy} 
\affiliation{University of Tsukuba, Tsukuba, Ibaraki 305, Japan} 
\affiliation{Tufts University, Medford, Massachusetts 02155} 
\affiliation{Waseda University, Tokyo 169, Japan} 
\affiliation{Wayne State University, Detroit, Michigan  48201} 
\affiliation{University of Wisconsin, Madison, Wisconsin 53706} 
\affiliation{Yale University, New Haven, Connecticut 06520} 
\author{T.~Aaltonen}
\affiliation{Division of High Energy Physics, Department of Physics, University of Helsinki and Helsinki Institute of Physics, FIN-00014, Helsinki, Finland}
\author{J.~Adelman}
\affiliation{Enrico Fermi Institute, University of Chicago, Chicago, Illinois 60637}
\author{T.~Akimoto}
\affiliation{University of Tsukuba, Tsukuba, Ibaraki 305, Japan}
\author{M.G.~Albrow}
\affiliation{Fermi National Accelerator Laboratory, Batavia, Illinois 60510}
\author{B.~\'{A}lvarez~Gonz\'{a}lez}
\affiliation{Instituto de Fisica de Cantabria, CSIC-University of Cantabria, 39005 Santander, Spain}
\author{S.~Amerio}
\affiliation{University of Padova, Istituto Nazionale di Fisica Nucleare, Sezione di Padova-Trento, I-35131 Padova, Italy}
\author{D.~Amidei}
\affiliation{University of Michigan, Ann Arbor, Michigan 48109}
\author{A.~Anastassov}
\affiliation{Rutgers University, Piscataway, New Jersey 08855}
\author{A.~Annovi}
\affiliation{Laboratori Nazionali di Frascati, Istituto Nazionale di Fisica Nucleare, I-00044 Frascati, Italy}
\author{J.~Antos}
\affiliation{Comenius University, 842 48 Bratislava, Slovakia; Institute of Experimental Physics, 040 01 Kosice, Slovakia}
% ...
\collaboration{CDF Collaboration\footnote{With visitors from $^a$University of Athens, 15784 Athens, Greece, 
$^b$Chinese Academy of Sciences, Beijing 100864, China, 
$^c$University of Bristol, Bristol BS8 1TL, United Kingdom, 
$^d$University Libre de Bruxelles, B-1050 Brussels, Belgium, 
$^e$University of California Irvine, Irvine, CA  92697, 
$^f$University of California Santa Cruz, Santa Cruz, CA  95064, 
$^g$Cornell University, Ithaca, NY  14853, 
$^h$University of Cyprus, Nicosia CY-1678, Cyprus, 
$^i$University College Dublin, Dublin 4, Ireland, 
$^j$University of Edinburgh, Edinburgh EH9 3JZ, United Kingdom, 
$^k$University of Heidelberg, D-69120 Heidelberg, Germany, 
$^l$Universidad Iberoamericana, Mexico D.F., Mexico, 
$^m$University of Manchester, Manchester M13 9PL, England, 
$^n$Nagasaki Institute of Applied Science, Nagasaki, Japan, 
$^o$University de Oviedo, E-33007 Oviedo, Spain, 
$^p$Queen Mary, University of London, London, E1 4NS, England, 
$^q$Texas Tech University, Lubbock, TX  79409, 
$^r$IFIC(CSIC-Universitat de Valencia), 46071 Valencia, Spain, 
}}
\noaffiliation


\ifprl
\author{}
\fi

\ifprl
\date{\today}
\else
\pagewiselinenumbers{}
\fi

\ifprl

\else
\begin{center}
  \begin{large}
{ \bf Search for Doubly Charged Higgs Bosons with Lepton-Flavor-Violating Decays involving Tau Leptons}
  \end{large}
\end{center}
\fi

\begin{abstract}
We search for pair production of doubly charged Higgs particles ($H^{\pm\pm}$) followed by decays into electron-tau ($e \tau$) and muon-tau ($\mu \tau$) pairs using data (350 pb$^{-1}$) collected from $\bar{p} p$ collisions at $\sqrt{s}=1.96$ TeV by the CDF II experiment. We search separately for cases where three or four final-state leptons are detected, and combine results for exclusive decays to left-handed $e\tau$ ($\mu \tau$) pairs. We set an \hpm\ lower mass limit of 114 (112)\gevcc\ at the 95\% confidence level.
\end{abstract}

\ifprl
\fi

\ifprl
\maketitle{}
\fi


The Standard Model (SM) Higgs mechanism provides a framework in which particles can acquire mass while preserving local gauge invariance. 
The complex scalar Higgs doublet of the SM is just one of many viable implementations, and many extensions to the SM contain Higgs triplets \cite{Cheng:1980qt,Mohapatra:1979ia,akeroyd-2005-72}.
For example the left-right symmetric ($SU(2)_{L} \times SU(2)_{R} \times U(1)_{B-L}$) extension of the electroweak force \cite{Mohapatra:1979ia} casts parity violation as a low-energy phenomenon by invoking a right-handed weak interaction broken above the electroweak scale.  This model predicts small but nonzero neutrino masses (consistent with recent experiments \cite{superk,sno}) related to the suppression of the right-handed weak current \cite{Mohapatra:1979ia}. 
Another model with an extended Higgs sector is the Higgs triplet model \cite{akeroyd-2005-72}, which predicts a massive left-handed Majorana neutrino without requiring a right-handed neutrino. An important phenomenological feature of the above models is the prediction of doubly charged Higgs bosons ($H^{\pm \pm}$) as part of a Higgs triplet. 
Doubly charged Higgs bosons couple to Higgs and electroweak gauge bosons and either left-handed or right-handed charged leptons ($\ell$), and are respectively denoted $H_L^{\pm \pm}$ or $H_R^{\pm \pm}$ \cite{noR}.

The only significant production mode at the Fermilab Tevatron is predicted to be $q\bar{q}$$\ra$$\gamma^* /Z$$\ra$$H^{++}H^{--}$, and the leptonic decay modes dominate for $\hpm$ in the mass range $m(\hpm)$$<$$(m(W^{\pm})+m(H^\pm))$\cite{Muhlleitner:2003me}. Lepton-flavor-violating (LFV) decay modes are allowed, and may be particularly large (e.g., the branching fraction for the $\mu\tau$ mode may be near 1/3) in the Higgs triplet model if the mass hierarchy of the quarks and charged leptons also holds for the neutrino sector\cite{ma-2000-85}.  

The $\hpml$ ( $\hpmr$) is excluded below 99$\gevcc$ (97$\gevcc$) at the 95\% C.L. by previous searches at LEP \cite{leplimits}, assuming production cross sections according to the left-right symmetric models \cite{Mohapatra:1979ia} and 100\% branching ratio to any one dilepton decay channel. Recent searches from the Fermilab Tevatron have resulted in 95\% C.L. lower mass limits of 136, 133, and 115$\gevcc$ for $\hpml$ in the $\mu\mu$, $ee$, and $e \mu$ channels, respectively, and a lower mass limit of 113$\gevcc$ for the $\hpmr$ in the $\mu \mu$ channel \cite{Acosta:2004uj}.

We present the first results from hadron colliders on $H^{++}_L H^{--}_L$ pair production and subsequent decay through LFV channels involving taus.  We use data corresponding to an integrated proton-antiproton luminosity of $\approx$350 pb$^{-1}$ \cite{lumi} collected at $\sqrt{s}=1.96$ TeV by the CDF II experiment at the Fermilab Tevatron, and set mass limits in the left-right symmetric model \cite{Mohapatra:1979ia,Muhlleitner:2003me} for exclusive decays in the $e\tau$ and $\mu\tau$ channels. We present limits on the cross section times branching ratio squared, $\sigma\times  {B}^2$, which can be interpreted in the context of various models \cite{Muhlleitner:2003me}.

CDF II \cite{cdfprd,more}, a cylindrical detector with concentric layers, has inner silicon strip detectors (SVX) and a wire drift chamber (COT) for tracking inside a solenoidal coil. The COT provides tracking in the pseudorapidity region $|\eta|$$\lesssim$1.3, while the SVX covers the region $|\eta|$$\lesssim$1.9. At radii outside the solenoid coil, sampling electromagnetic and hadronic calorimeters cover the region $|\eta|$$<$3.6 with a projective tower geometry. In the central region ($|\eta|$$\le$ 1.0), the electromagnetic calorimeter (CEM) has an embedded multi-wire proportional chamber (CES), with anode wires parallel to the beam direction, and orthogonal cathode strips. The CES has 2 cm strip/wire spacing and provides $\approx$ 2 mm spatial resolution of electromagnetic showers. The region 1.1$\le$$|\eta|$$\le$3.6 is covered by the ``plug'' electromagnetic (PEM) and hadronic calorimeters.  At the largest radii there are scintillator and drift tube muon detectors in the region $|\eta|$$<$1.5.

We use several sets of selection criteria to characterize lepton candidates. All ``tight'' leptons must be in the central region, while ``loose'' leptons satisfy $|\eta|$$<$1.3.   
Tight electrons \cite{zttprd} have tracks in the COT matched to energy clusters in both the CEM and CES.  They pass requirements on the electromagnetic to hadronic calorimeter deposition ratio, the CEM energy to COT track momentum ratio, and a tower-to-tower energy sharing variable.
Loose electrons only have tracks matched to CEM or PEM clusters with electromagnetic to hadronic calorimeter deposition ratios consistent with the electron hypothesis. Tight muons \cite{Acosta:2004uj} are minimally ionizing in the calorimeters and have tracks in the COT that extrapolate to hits in the outer muon detectors. Loose muons are simply isolated tracks, as described below.
In order to suppress background from jets misidentified as leptons, an electron or muon is selected to be isolated by requiring that the sum of the transverse momenta of all other tracks in a cone of angle 0.4 radians with respect to the lepton's direction be less than 2\gevc.

Identification of hadronically decaying taus ($\tau_h$) is fully described elsewhere \cite{zttprd}.  In tau reconstruction, all tracks are assumed to correspond to charged pions, and all trackless CES/CEM clusters are assumed to correspond to \piz\ mesons.   A tight $\tau_h$ must have 1 or 3 localized tracks, and can have additional localized  \piz\ candidates. The localization is defined by a variable size ``signal cone'' (between 3$^\circ$ and 10$^\circ$, depending on the tau's momentum) around the highest $p_T$ track associated with the $\tau_h$. The region between the signal cone and a larger 30$^\circ$ cone 
serves as an isolation annulus in which the summed $p_T$ of all tracks must be less than 2\gevc\ and the summed $E_T$ of all \piz\ mesons must be less than 0.5\gev. The 4-momentum of a $\tau_h$ is taken to be the vector sum of the 4-momenta of the tau's tracks and \piz\ candidates in the signal cone. The charge of a $\tau_h$ is the sum of the charges of its tracks, and must equal $\pm 1$.  
A loose $\tau_h$ is the same as a tight $\tau_h$ in the region $|\eta|$$<$1.0, but has additional acceptance for 1.0$<$$|\eta|$$<$1.3.  
Since the CES does not cover the latter region, \piz\ related cuts are dropped, and the energy of a loose $\tau_h$ is estimated from the plug calorimeters.

To increase signal acceptance, systems of one or three isolated, localized tracks in the region $|\eta|$$<$1.3 are also considered as loose lepton candidates. For such candidates, the signal and isolation cone sizes are 10$^\circ$ and 30$^\circ$ respectively. These ``isolated track systems'' (ITSs) have acceptance for $e$, $\mu$, and $\tau$ leptons.  
The efficiencies of lepton reconstruction, identification, and isolation requirements are measured in data using electrons from decays of $\Upsilon$ mesons, electrons and muons from decays of $Z$ bosons, and taus from $W$ bosons.
  
We require at least three reconstructed isolated charged leptons to suppress large cross-section backgrounds such as dijets, $\gamma$ + jets, and $W(\ra \ell\nu_\ell)$+ jets. Events are classified according to the number of  isolated high $p_T$ leptons detected, and separate selections are used for the 3-$\ell$ and 4-$\ell$ signatures. 
The data are collected by lepton plus isolated track triggers \cite{Anastassov:2003vc}. 
These triggers require one central lepton ($e$ or $\mu$) and a second central isolated track. 
The integrated luminosities of the $e \tau$ and $\mu \tau$ samples are 350 pb$^{-1}$ and 322 pb$^{-1}$, respectively. Trigger efficiencies for electrons (muons) are estimated from events with photon conversions and $Z$$\ra$$ee$ ($J/\psi$$\ra$$\mu \mu$ and $Z$$\ra$$\mu \mu$) decays. 
The efficiency for the isolated track is measured from a jet sample.  The overall trigger efficiencies are $\approx$ 95\% for $\hpm$ masses in the range 80-130\gevcc.
The specific lepton requirements for the $e\tau$ and $\mu \tau$ searches are summarized in Table {\rm \ref{leptons.table}}. 

We use CTEQ5L parton density functions (PDFs) in the {\sc pythia} generator \cite{pythia} and a {\sc geant}-based \cite{geant} detector simulation, scaled to next-to-leading order (NLO) cross sections \cite{Muhlleitner:2003me}, to estimate the signal and background processes. Our signal MC samples scan the $\hpm$ mass range 80-130$\gevcc$ at $10\gevcc$ intervals.
The potential SM backgrounds for both the 3-$\ell$ and 4-$\ell$ searches are: $Z$/$\gamma^*$$\ra$ leptons produced in association with $\ge$1 hadronic jet(s) or photon(s); $ZZ$ and $WZ$ with both bosons decaying leptonically; $t\bar{t}$  with leptonically decaying $W$ bosons;  $W$ bosons decaying leptonically produced in association with $\ge$2  hadronic jets; and ``QCD'' events (no leptons, $\ge$3 hadronic jets).  For the $e \tau$ signature, $\gamma$ + hadronic jets events are also a potential background, while cosmic ray muons are a potential background in the $\mu \tau$ channel.  The backgrounds with the larger production cross sections (e.g., QCD, $W$) are suppressed by multiple powers of the lepton misidentification rates ($\approx$10$^{-2}$ for jet$\ra$$\tau$, and $\approx$10$^{-4}$ for jet$\ra$$e,\mu$).           

 Event selection for the 3-$\ell$ events begins with the removal of events that are consistent with cosmic ray muons \cite{cosmic} or low-mass Drell-Yan lepton pairs ($M(e^+e^-)$$<$30$\gevcc$; $M(\mu^+\mu^-)$$<$$30\gevcc$).  
Also, events consistent with $Z+\gamma$ production with the photon misidentified as an electron are efficiently removed by requiring at least 20\gev\ of missing transverse energy ($\met$) \cite{more}.  
Signal events with at least one $\tau$ decaying to an electron typically have $\met$ $>$20$\gev$, due to the significant fraction of the $\tau$'s energy carried off by the two neutrinos, while $Z+\gamma$ events are typically well measured, and thus have small \met.  
Similarly, in the 4-$\ell$ search, events consistent with having four final-state electrons must have at least 20\gev\ of \met.  No attempt is made to reconstruct the full $\hpm$ mass, but we do require the presence of a like-sign $e\tau$ or $\mu \tau$ pair with an invariant mass in the range 30-125\gevcc. This selection is nearly 100\% efficient for signal but reduces diboson and top backgrounds.  

To further reduce backgrounds, in particular $Z$ + jets, we impose a requirement on the scalar sum of the lepton transverse energies and \met ($Y_T$). The $Y_T$ requirement depends on whether an event is tagged as a $Z$ boson decay.  It is more efficient to remove events consistent with $Z$ boson decays by $Y_T$ than by a direct mass cut, because some of the signal has oppositely charged leptons in the $Z$ mass range, but large $Y_T$ values compared to $Z$ + jets events.
The $Y_T$ cut values for tagged and untagged events, as well as the mass window used in $Z$ boson tagging, are optimized by running pseudoexperiments and choosing the sets of cut values that result in the best expected limits on $H^{++}$.    The $e\tau$ search uses $Y_T$ cuts of 190\gev\ for untagged events and  300\gev\ for events tagged as $Z$ boson candidates, defined as an $e^+ e^-$ pair in the mass range 71-111\gevcc.  The $\mu\tau$ search uses $Y_T$ cuts of 190\gev\ for untagged events, and 350\gev\ for events tagged as $Z$ boson candidates, defined as a $\mu^+\mu^-$ pair in the mass range 76-116\gevcc.  
In the $\mu \tau$ analysis, a muon with a severely mismeasured $p_T$ may lead to spuriously high $Y_T$.  
We minimize the mismeasurement risk by imposing additional cuts on the highest $p_T$ tracks in the events.  

Events with four isolated leptons have less background than trilepton events, so less restrictive cuts are applied.  We first require $Y_T$$>$$120\gev$.  Events tagged as $Z$ bosons are required to have $\met$$>$$20\gev$ in the $e\tau$ search and $Y_T$$>$$150\gev$ in the $\mu\tau$ search.  As with $Y_T$ and $Z$-veto for the 3-$\ell$ channels, pseudoexperiments were conducted with various values of both cuts, and the cuts that resulted in the best expected limits  were chosen for each analysis. The acceptances for the 3-$\ell$ and 4-$\ell$ channels are roughly equal, and the combined  acceptance grows approximately linearly with $H^{\pm \pm}$ mass from 8\% at $85\gevcc$ to 14\% at $135\gevcc$.  Observed and expected event yields for signal and background for the 3-$\ell$ and 4-$\ell$ searches are shown in Table \ref{tally.table}. The signal event yields assume $\sigma$$\times$$B^2=89.4 fb$, corresponding to exclusive decays of 110 \gevcc\ \hpm\ to $e \tau$ ($\mu \tau$) pairs in models \cite{Mohapatra:1979ia} and \cite{akeroyd-2005-72}.  The $Z$ + jets process is the most significant single background, with  0.15$^{+.11}_{-.07} (stat)$ expected events for each of the combined (3-$\ell$ + 4-$\ell$) $\mu \tau$ and $e \tau$ searches.  The combined background from $WZ$ and $ZZ$ production amounts to 0.12$\pm$0.02 (0.20$\pm$0.02) events for the $e \tau$ ($\mu \tau$) search.   $t\bar{t}$ background is 0.01$^{+0.02}_{-0.01}$ (0.06$^{+0.02}_{-0.01}$) events in the $e \tau$ ($\mu \tau$) search.  Cosmic ray, $\gamma$ + jets, and QCD backgrounds are negligible and determined from data. 

 Systematic uncertainties on backgrounds from NLO cross section uncertainties are 4\% for $Z$ and $W$ boson production processes and 8\% for diboson and top quark production processes \cite{Campbell:1999ah}.  A 6\% uncertainty applies to the integrated luminosity of our dataset.  A 28\% (21\%) systematic uncertainty is used for the $W\ra$$\ell \nu_\ell$  ($Z\ra$$\ell\ell$) background predictions to account for imperfect knowledge of the jet$\ra$$\tau_h$ misidentification rate.  Imperfect simulation of the track curvature resolution is accounted for by a 0.1 event systematic uncertainty on the combined backgrounds for the $\mu\tau$ search. The combined systematic uncertainty for all backgrounds amounts to 0.04 (0.11) events for the $e\tau$ ($\mu\tau$) search.  The total uncertainties on backgrounds, shown in Table \ref{tally.table}, are statistically dominated.  
 Systematic uncertainties on the signal cross section include NLO cross section uncertainties (7.5\%) \cite{Muhlleitner:2003me}, luminosity (6\%) \cite{lumi}, and parton density function (PDF) uncertainty (5\%) \cite{pdf}. The uncertainty on signal acceptance (6.1\%) is driven by uncertainties on track isolation efficiency (4.5\% and 6\% for 3-$\ell$ and 4-$\ell$ channels, respectively), and \piz\ isolation efficiencies (1.5\% and 2\% for 3-$\ell$ and 4-$\ell$ channels, respectively).
  
\begin{figure}
  \begin{center}
    \includegraphics[width=8.2cm,clip=]{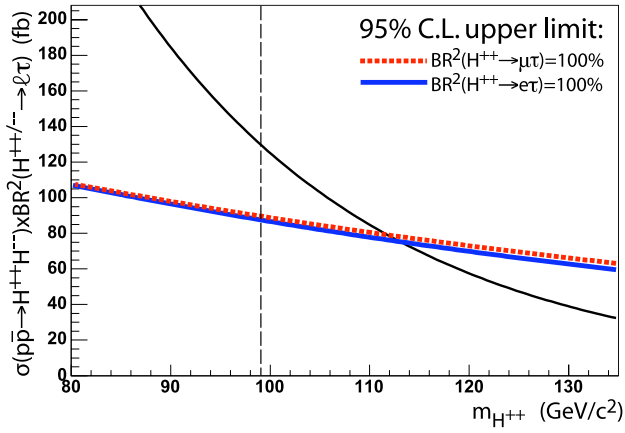}
    \caption{ Theoretical production cross sections for the pair production of left-handed $\hpm$, and 95\% C.L. limit curves for $\sigma (p\bar{p} \ra H^{++}H^{--}\rightarrow) \times B^2( \ell^+\tau^+ \ell^- \tau^-)$, for $\ell=e$(solid), $\mu$(dashed). The vertical dashed line corresponds to limits from experiments at LEP2 for exclusive \hpml decays to any one dilepton channel \cite{leplimits}.}
\label{danewlimits}
  \end{center}
\end{figure}

 We find that the background predictions agree with data in all control samples, including samples in the kinematic region $Y_T$$<$150$\gev$ enriched with QCD, $Z$ boson, and $W$ boson events.  To check our predictions in the high-$Y_T$ regime while keeping the analysis ``blind,'' we check the number of events that pass all analysis selections except track isolation for the second tight lepton (Table I). After finalizing all selection requirements and our limit setting procedure, we search the signal regions in both the 3-$\ell$ and 4-$\ell$ channels. We observe no events in either the 3-$\ell$ or 4-$\ell$ channels for both the $\mu \tau$ and $e \tau$ searches, which is consistent with the SM backgrounds of 0.24$^{+0.27}_{-0.24}$ $e \tau$ events and 0.39$\pm$0.23 $\mu \tau$ events.  Limits are set using a Bayesian method based on a Poisson likelihood, with a flat prior for signal cross section and Gaussian priors for uncertainties on signal, background acceptance, and integrated luminosity. The 3-$\ell$ and 4-$\ell$ channels are treated as separate measurements, taking into account correlated systematic uncertainties \cite{zttprd}.  
We set an upper $\sigma$$\times$$B^2$ limit for the process  $p\bar{p}$$\ra$$H_{L}^{++}H_L^{--}$$\rightarrow$$e^+\tau^+ e^- \tau^-$ of 74 fb at the 95\% C.L., which corresponds in models \cite{Mohapatra:1979ia} and \cite{akeroyd-2005-72} to a mass limit of 114\gevcc. The process $p\bar{p}$$\ra$$H_{L}^{++}H_L^{--}$$\rightarrow$$\mu^+\tau^+ \mu^- \tau^-$ is excluded above a cross section of 78 fb at the 95\% C.L., corresponding to a mass limit of 112\gevcc\ in the same models. 
The exclusion curves are shown in Fig. \ref{danewlimits}.

We thank the Fermilab staff and the technical staffs of the participating institutions for their vital contributions. This CDF work was supported by the U.S. Department of Energy and National Science Foundation; the Italian Istituto Nazionale di Fisica Nucleare; the Ministry of Education, Culture, Sports, Science and Technology of Japan; the Natural Sciences and Engineering Research Council of Canada; the National Science Council of the Republic of China; the Swiss National Science Foundation; the A.P. Sloan Foundation; the Bundesministerium f\"ur Bildung und Forschung, Germany; the Korean Science and Engineering Foundation and the Korean Research Foundation; the Science and Technology Facilities Council and the Royal Society, UK; 
the Institut National de Physique Nucleaire et Physique des Particules/CNRS; the Russian Foundation for Basic Research; the Comisi\'on Interministerial de Ciencia y Tecnolog\'{\i}a, Spain; the European Community's Human Potential Programme under contract HPRN-CT-2002-00292; and the Academy of Finland.

\begin{table}[h]
\begin{center}
\begin{small}
\begin{tabular}{ccccc}
\hline
\hline

Signature    &Lepton                                      &Flavor     &  $E_T$ ($P_T$) & $|\eta|$ \\
\hline
\multirow{3}{*}{3-$\ell$}& 1$^{st}$ (tight) &$e$& $>20\gev$ & $<1.0$\\ 
&2$^{nd}$ (tight)&  $\tau_h$ or $e$& $>15\gev$&$<1.0$\\
&3$^{rd}$ (loose)&  $\tau_h$ or $e$& $>10\gev$& $<1.3$\\
\hline
4-$\ell$ &4$^{th}$ (loose)& Isolated Track&$>10\gevc$& $<1.3$\\
\hline
\hline
\end{tabular}
\end{small}
\caption{Kinematic and geometric lepton requirements (cut values) for the $e\tau$ search.  For the $\mu\tau$ search, the first lepton changes from $e$ to $\mu$, and the third lepton changes from $\tau_h$ or $e$ to isolated track. }
\label{leptons.table}
\end{center}
\end{table}

\begin{table}[h]
\begin{center}
\begin{tabular}{ccccc}
\hline
\hline
$e\tau$ &Selection& Exp. Signal &Background & Data\\
\hline
\multirow{4}{*}{3-$\ell$}&Lepton ID&   2.94$\pm$0.11&       37.8$\pm$1.3      &  34      \\
&$M_{LS}, M_{OS} $   &  2.89$\pm$0.11      &  35.4$\pm$1.2   &  29       \\
& $Y_T$/$Z$ veto         &  2.4$\pm$0.09        &  9.65$\pm$0.66   &   8        \\
&$Y_T$                         & 1.97$\pm$0.08         & 0.24$^{+0.27}_{-0.24}$    &  0      \\
\hline
\multirow{2}{*}{4-$\ell$}&Lepton ID& 1.61$\pm$0.07& 0.18$\pm$0.06  & 0 \\
& $Y_T$/$Z$ veto & 1.60$\pm$0.07 &0.04$^{+0.05}_{-0.04}$&0\\
\hline

$\mu\tau$&Selection&Exp. Signal&Background& Data\\
\hline
\multirow{4}{*}{3-$\ell$}&Lepton ID &  3.06$\pm$0.04      &30.0$\pm1.4$     &28\\
&$M_{LS}, M_{OS}$&2.99$\pm$0.04        &24.6$\pm$1.26     &20\\ 
&$Y_T$/$Z$ veto                    & 2.35$\pm$0.04         & 6.6$\pm$0.86       &7\\
&$Y_T$                    & 1.80$\pm$0.03         & 0.27$\pm$0.22   &0\\
\hline
\multirow{2}{*}{4-$\ell$}&Lepton ID&  1.65$\pm$0.03& 0.25$\pm$0.08  &  0 \\
&$Y_T$/$Z$ veto&       1.64$\pm$0.03& 0.14$\pm$0.05&0\\
\hline
\hline
\end{tabular}
\caption{Cumulative effect of selection requirements on signal (110\gevcc, $\sigma\times  {B}^2=89.4 fb$ ) and background in the 3-$\ell$ and 4-$\ell$ searches. $M_{LS}$ ($M_{OS}$) represent the invariant mass requirements on the like (opposite) sign leptons. The $Z$ veto refers to the additional $Y_T$ requirement on $Z$ boson tagged events.  The uncertainties are combined statistical and systematic.}
\label{tally.table}
\end{center}
\end{table}


\bibliography{PRL_OFFICIAL}

\begin{thebibliography}{20}
\bibitem[{noR()}]{noR}
\bibinfo{note}{With the present dataset, expected mass limits for
  $H_R^{\pm\pm}$ are much lower than the LEP2 limits because the $H_R^{\pm\pm}$
  pair production cross section is about half as large as that for
  $H_L^{\pm\pm}$.}
\bibitem[{mor()}]{more}
\bibinfo{note}{CDF uses a cylindrical coordinate system in which $\phi$ is the
  azimuthal angle, $\theta$ is the polar angle, $r$ is the radius from the
  nominal beamline, and $+z$ points from the nominal interaction point along
  the proton beam. The pseudorapidity is defined $\eta = -\ln[tan(\theta/2)]$.
  Calorimeter energy (track momentum) measured transverse to the beam is
  denoted as $E_T$ ($p_T$), and the total calorimetric transverse energy
  imbalance is denoted as $\met$.}
\bibitem[{pdf()}]{pdf}
\bibinfo{note}{We calculate this uncertainty from the changes in cross section
  due to $1\sigma$ variations in the eigenvectors in the CTEQ5M PDF}.
\end{thebibliography}
\end{document}